\documentclass[12pt, letterpaper, fleqn]{article}

\usepackage[magyar]{babel}
\usepackage{titlesec}
\usepackage{amsmath}
\usepackage{amsfonts}
\usepackage{fancyhdr}
\usepackage{enumitem}
\usepackage{xpatch}
\usepackage{amssymb}
\usepackage{amssymb}
\usepackage{amsmath}
\usepackage{amscd}
\usepackage{enumerate}
\usepackage{array}
\usepackage{tabularx}
\usepackage{t1enc}
\usepackage[mathscr]{eucal}
\usepackage{graphicx}
\usepackage[utf8]{inputenc}

\begin{document}

\centerline{\bf Quantum entanglement approaching with concurrence}

\centerline{\bf in the presence of chaos}

\vspace{0.5cm}

\centerline{A. F\"ul\"op} 

\vspace{0.3cm}

\centerline{ELTE E\"otv\"os Lor\'and University, Budapest, Hungary}

\vspace{1.0cm}

{\bf Abstract}

The concept of concurrence is researched to characterize the dynamical behavior of the bipartite systems. The quantum kicked top model has great significance in the qubit systems and the chaotic properties of the entanglement.
The eigenvalues of the reduced symmetric density matrix are determined, it allows us to understand this driven system to distinguish between regularity and chaoticity dynamics in the finite simulation, which depend on the strength excitation in the framework of the concurrence.

\section{\bf Introduction}

In this article we study the concurrence of the entanglement through the quantum kicked top model comparing with the spin coherent states, Dick quantum states and EPR correlated systems. 
The idea of the concurrence was introduced by the entanglement of formation describing the bipartite mixed entanglement system \cite{ap3}.  

The importance of the kicked top model is demonstrated by the large
number  of papers  \cite{csagj,bs,rlp2,lm2,bl1,mgtg,fmt1,7pgfhhw,10sgbcs,5mkjmfh,ac,af,13bs2} as the quantum simulation and the describing  the quantum entanglement.
The dynamic of quantum kicked top system (QKT) is introduced by Hamiltonian and Flouquet map. The chaoticity is discussed by the concurrence depending on the evaluation \cite{6kz,mkrsfh4,ap3}.

 The entanglement plays significant role in the operability and stability of the quantum computer resulting in the minimized effect.
 Between the quantum processor and the environment, the entanglement causes the noise and the computational errors. 
This is implemented by reliable quantum computing, it is important to understand and control the noise in quantum protocols \cite{4cmjpprp,5cmjppwhz,6phsdls,7ggcgbgc,8drgbgc,lm2,34sbdls,7a}.

 Even when a quantum processor is ideally isolated from the environment, i.e.,
in situations where the decoherence time of the processor is very large as compared to the computational time scales, the operability of the quantum computer is not yet guaranteed \cite{10bgdls,58skmalvs}.
Indeed, the presence of device imperfections hinders the implementation of any quantum computation.

A quantum computer corresponds to a quantum many-body system.
The interaction between the qubits has a remarkable role that makes up the computer's quantum registers to create the required amount of entanglement.
Small inaccuracies in the coupling constants cause errors in the operation of the device. Above a certain imperfection strength threshold (chaos limit), quantum chaos occurs \cite{10bgdls,11vvf,12gbgcdls,13gpbfbfmivit,11vmcarsg,35}.

In such a system, exponentially many states of the computational base are mixing after a chaotic time scale.
It gives an upper time limit to the stability of the superposition of its states encoded in the wave function of the quantum computer.
For quantum computers to function properly and meet error calculation schemes, then many gates must be used within the chaotic time scale.

Structure of the article:
In the section (\ref{ent}), we introduce the idea of entanglement and concurrence in a bipartite Hilbert space. In the next section (\ref{dm}), discuss the form of the density matrix by the pairwise method. Then some multiqubit system is characterized as the spin coherent system, Dick states and the EPR-correlated ensembles considering the concurrence and entanglement (\ref{pms}).
 After these cases, we study the quantum Kicked Top model (\ref{ktm}) by a three-qubit system. The concurrence is determined depending on the time. 
The chaoticity is shown in this system.  This periodic driven system as a function the strange of excitation $\kappa$ is considered for different values of strange driven. 

\section{Entanglement}\label{ent}

The pure state $|\Psi>$ is separable if a bipartition Hilbert space $H=H_A\otimes H_B$ i.e. $|\psi>=|a>\otimes |b>$, where $|a>$ and $|b>$ vectors being an element of Hilbert subspace  $H_A$ and $H_B$.  The pure state $\psi$ is entangled if it is not separable.

The pure bipartite entanglement can be written by Von Neumann entropy and concurrence. For the mixed states, the Von Neumann entropy does not give a good solution \cite{38wkw} in some cases. 
The entanglement of formation $E_F$  \cite{39chbdpdjaswkw}
describes the mixed states of the bipartite entanglement \cite{cmmgbggcgc}, where the mixed states can be written with density matrix $\rho=\sum_ip_i |\psi_i><\psi_i|$ \cite{39chbdpdjaswkw}. The expression $E_F(\rho)$ means the average entanglement of the pure states at a given decomposition of the mixed state $\rho$. Let us minimize overall possible decompositions:

\begin{eqnarray}\label{10ef}
E_F(\rho)=\min_{\{p_i, \Psi_i\}}\sum_i p_i E(|\psi_i>)
\end{eqnarray}
where $E(|\Psi_i>)$ is the amount of entanglement of the pure state $|\Psi_i>$.

\paragraph{Concurrence}
This expression (\ref{10ef}) can be rewritten on two-qubit system,  the concurrence $C$ \cite{40wkw} is  defined by the expression $E_F$:
\begin{eqnarray}
E_F(\rho)=h\left(\frac{1}{2}\left[1+\sqrt{1-C(\rho)^2}\right]\right)
\end{eqnarray}
where $h(x)$ means a binary entropy function which is determined following
\begin{eqnarray}
h(x)=-x\log_2x-(1-x) \log_2(1-x)
\end{eqnarray}
The concurrence $C(\rho)$ of the two-qubit state $\rho$ is introduced:
\begin{eqnarray}\label{con}
C(\rho)=\max\{0,c_\lambda\}
\end{eqnarray}
where $c_\lambda=\lambda_1-\lambda_2-\lambda_3-\lambda_4$, the $\{\lambda_i\}$ being the square roots of the eigenvalues of the matrix $\rho(\sigma_y\otimes \sigma_y)\rho^*(\sigma_y \otimes \sigma_y)$ in decreasing order and $\sigma_y$ a Pauli matrix. In this  definition of the complex conjugation the basis is $\{|00>,|01>,|10>,|11>\}$, where $|0>$, $|1>$ are the eigenstates of the $\sigma_z$. We can see that $E_F$ depends monotonously \cite{gv} on the concurrence, which changes between 0 for separable states and 1 for maximally entangled states and the $c_\lambda$ take values in $[-1/2,1]$. Therefore the state is separable if $c_\lambda\le 0$ and entangled otherwise.

The multiparticle entanglement is more complex. Different measures of multipartite entanglement have been given different results qualitatively  \cite{44db}.

\section{Density matrix  on bipartite system}\label{dm}

This can be considered whether $N$ quantum systems have an entangled state or it doesn't. The main behavior of the reduced density matrix can be shown on a smaller bipartite system. In the case of a pure state in the two-particle model, the eigenvalues $t_1$ of the reduced density matrix are studied in both systems.

Then this form $E=-\sum_i t_i \log_2 t_i$ which means the asymptotic ratio between $n$ and $m$, where $n$ is the number of pairs in the appropriate state extracted $m$ maximally entangled pairs of states \cite{xwkm}.

The mixed state can be defined by a density matrix $\rho_{12}$ that is similar to its minimum value of the weighted average of $E$ over the wave functions, where a
two-particle density matrices can be introduced in the form of a weighted sum.

Because the  $\rho_{12}$ can be given in several ways as a weighted sum of pure state projections, the minimal value is chosen as the appropriate value. 
In this system the entanglement can be described by the non-positive value of the partially transposed density matrix \cite{25ap}. Then this quantity is defined by the expression (\ref{10ef}) in the section (\ref{ent}).
The concurrence $C$ is introduced by the term (\ref{con}) according to the entanglement (\ref{10ef}), where the quantities $\lambda_i$ are the square roots of the eigenvalues of the matrix product $\varrho_{12}=\rho_{12}(\sigma_{1y}\otimes\sigma_{2y})\rho^*_{12}(\sigma_{1y}\otimes\sigma_{2y})$ in decreasing order, where the $\rho_{12}^*$ means the complex conjugation of $\rho_{12}$, and $\sigma_{iy}$ are Pauli matrices. The eigenvalues of $\rho_{12}$ are real and non-negative and $\rho_{12}$ is not necessarily Hermitian. The range of the concurrence changing between zero and unity i.e. from an unentangled state to a maximally entangled state.

\paragraph{Pairwise method}

If the system satisfies the symmetry condition, i.e. the density matrix  is symmetric during the change of the system \cite{xwkm}, then the expression is following
\begin{eqnarray}\label{36}
\rho_{12}=\left( 
\begin{array}{cccc}
v_+ & x^*_+ & x_+^* & u^* \\
x_+ & w & y^* & x^*_{-}\\
x_+ & y & w & x^*_{-}\\
u & x_{-} & x_{-} &v_{-}
\end{array}
  \right)
\end{eqnarray}
where the basis of the matrix elements are
$\{|00>,|01>,|10>,|11>\}$.
These are written by the  expectation
values of Pauli spin matrices in the following:
\begin{eqnarray}\label{37}
\begin{array}{rcl}
v_{\pm}&=&\frac{1}{4}(1\pm 2<\sigma_{1z}>+<\sigma_{1z}\sigma_{2z}>),\hspace{0.5cm} y =<\sigma_{1+}\sigma_{2-}>\\
x_{\pm}& =& \frac{1}{2}(<\sigma_{1+}>\pm <\sigma_{1+}\sigma_{2z}>),\hspace{1.3cm} w = \frac{1}{4}(1-<\sigma_{1z}\sigma_{2z}>), \\
u& =& \frac{1}{4}(<\sigma_{1x\sigma_{2x}}>-<\sigma_{1y}\sigma_{2y}>+i2<\sigma_{1x}\sigma_{2y}>).\\
\end{array}
\end{eqnarray}
Then the states of two-qubits are studied, they arise from a symmetric multi-qubit state system. First of all, the symmetric pure states are discussed, and then the state of the $N$-qubit system can be described by the orthonormal basis $|S,M>$ $M=-S,-S+1,\dots S$, where $S = N/2$. The $|S,M>$ states are the symmetric Dicke states, they have the eigenstates of the collective spin operators  $\overrightarrow{S}^2$ and $S_z$:
\begin{eqnarray}\label{38}
S_{\alpha}=\frac{1}{2}\sum_{i=1}^N \sigma_{i\alpha}, \;\; \alpha=x,y,z.
\end{eqnarray}
 The form of the number operator  can be specified
$\mathcal{N} = S_z + N/2$ and the state of the number following
\begin{eqnarray}\label{39}
\begin{array}{rclrrcl}
|n>_N &\equiv& |N/2,-N/2+n>_N, & \mathcal{N}|n>_N&=&n|n>_N
\end{array}
\end{eqnarray}
which plays an important role later on. The number of qubits in the state $|0>$ represents the eigenvalue $n$ of the number operator $\mathcal{N}$, where  the states $|0>_N$  and  $|1>_N$ can  be formed as
$|0>_N = |111\dots 1>$ and
$|1>_N =\frac{1}{\sqrt{N}}(|0111\dots 1>+|101\dots 1>+ \dots +|111\dots 0>)$.
The state $|1>_N$ is  referred as an $N$-qubit $W$ state \cite{1wdgvjic,20xw}.
 The Pauli matrices are expressed  as the multi-qubit state
due to the symmetry condition in the following form
\begin{eqnarray}\label{42}
\begin{array}{rclrrcl}
<\sigma_{1\alpha}>& =& \frac{2<S_\alpha>}{N},&\hspace{1.0cm}<\sigma_{1+}>& =& \frac{<S_+>}{N},\\
<\sigma_{1\alpha}\sigma_{2\alpha}>& =& \frac{4<S^2_{\alpha}>-N}{N(N-1)}&\hspace{1.0cm}<\sigma_{1x}\sigma_{2y}>& =& \frac{2<[S_{x},S_y]_+>}{N(N-1)}\\
<\sigma_{1+}\sigma_{2z}>& =& \frac{<[S_+,S_z]_+>}{N(N-1)}&\\
\end{array}
\end{eqnarray}
where $[A,B]_+$ denotes the anticommutator for operators. The density matrix elements of $\rho_{12}$ means as their expected values of the collective operators
\begin{eqnarray}\label{43}
\begin{array}{rclrcl}
v_{\pm}& =& \frac{N^2-2N+4<S_z^2>\pm 4<S_z>(N-1)}{4N(N-1)},& \hspace{0.5cm}
x_{\pm} & =& \frac{(N-1)<S_+>\pm <[S_+,S_z]_+>}{2N(N-1)}\\
w & =&  \frac{N^2-4<S_z^2>}{4N(N-1)},& \hspace{0.5cm} y& =& \frac{2<S_x^2+S_y^2>-N}{2N(N-1)}\\
u&=& \frac{<S_x^2-S_y^2>+i<[S_x,S_y]_+>}{N(N-1)}=\frac{<S_+^2>}{N(N-1)}.\\
\end{array}
\end{eqnarray}

\section{Basic multiqubit states}\label{pms}

In this section, we study such cases where $N$ two-values systems are written by a pure state that is invariant to permutations of particles \cite{xwkm}.

\paragraph{Spin coherent states}

This is a separable state. Because the rotation of the spin state $|S,M=S>$ produces the coherent state of the spin, which corresponds to the product state of all $N$ particles in the $|0>$ state.  Applying the above procedure, the
expression of spin coherent states is achieved:
\begin{eqnarray}\label{44}
|\eta>=(1+|\eta|^2)^{-N/2}\sum_{n=0}^N\left( \begin{array}{c}
N\\
n
\end{array}\right)^{1/2}\eta^n |n>_N,
\end{eqnarray}
where $\eta$ is a real value. The result of this calculation is the following
$v_+=\eta^4$, $x^*_+=x_+=\eta^3$, $w=y=u=u^*=\eta^2$, $x_-=x^*_-=\eta$, $v_-=1$.
This corresponds to the two-particle state which is produced by the product state of two rotated 1/2 spin particles in the state $(\eta |0>+|1>)/\sqrt{1+\eta^2}$.
The matrix product $\rho_{12}$ is a zero matrix of $4\times4$, which shows the role of $\sigma_y$ Pauli matrices. The expression $\rho_{12}$ means the projection operator in the spin states in which the direction is defined in the $xz$ plane, applied $\sigma_y$ satisfies a $180^\circ$ rotation in the $xz$ plane, so $\rho_{12}$ the vanishing product of projection operators on two orthogonal subspaces.

 In this case there is the concurrence disappears $C=0$, i.e. there is no entanglement.

\paragraph{Dicke state}

\begin{figure}
\begin{center}
\includegraphics[width=6.0cm]{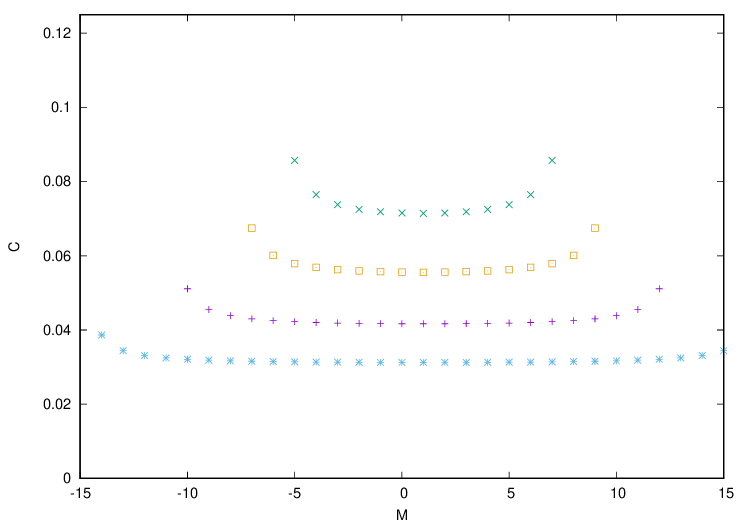}
\caption{ The concurrence in the Dicke state for different number
N  (N = 15, N = 20, N=25 and N = 30).}\label{fig-1}
\end{center}
\end{figure}

Dick states $|N/2,M>$, introduced as effective number state for a given number of particles, including the states  $|0>$ and $|1>$. Thus, due to rotating all the spins, a separable spin coherent state is created with a binomial distribution over the different Dicke states. Then  the reduced density matrix $\rho_{12}$ can be specified: $w=y$, $v_+,v_-$, the other elements $u=u^*=x_+=x^*_+=x_-=x^*_-=0$
where the matrix elements
\begin{eqnarray}\label{47}
\begin{array}{rclrcl}
v_{\pm}& =& \frac{(N\pm 2M)(N-2\pm2M)}{4N(N-1)},&
w & =& \frac{N^2-4M^2}{4N(N-1)}
\end{array}
\end{eqnarray}
The suitable concurrence of a simple density matrix of the expression (\ref{38}) is obtained by \cite{40wkw}. The form corresponding to the density matrix is
$C=2\max\{0,y-\sqrt{v_+v_-}\},$
where the form $2y+v_++v_-=1$ is applied. Considering the previous results the concurrence is following
\begin{eqnarray}\label{49}
C=\frac{1}{2N(N-1)}\left\{N^2-4M^2-\sqrt{(N^2-4M^2)[(N-2)^2-4M^2]}\right\} 
\end{eqnarray}
  The Figure (\ref{fig-1}) shows the values of $C$ depending on different $N$ and $M$. For any Dicke state, without those for which $|M|$ get a maximum value if two-particles are removed, they will be in an entangled state.  The concurrence takes a constant value near $M = 0$. Then the variation of $M$ is studied in a small range around $M = 0$, provided the initial binomial distribution which has the mean value, the concurrence will be very close to the result, $C = 1/(N - 1)$ for $M = 0$.
  
The concurrency of Dicke states $|N/2,M = \pm(N/2 - 1)>$ is $C = 2/N$.
These states are the same as a $W$ states that have the largest concurrence considering the symmetric states.

\paragraph{EPR-correlated ensembles}

 The EPR-correlated ensemble is studied which has great significance in theoretical quantum physics. This state is not invariant in the case of arbitrary permutation of particles, but only during permutations that exchange the particles within each ensemble: 
$(J_{1x} - J_{2x})|\psi>í = 0$ and 
$(J_{1y} + J_{2y})|\psi > = 0$.
This expression  can be written equivalently:
$(J_{1+} - J_{2-})|\psi > = 0$ and 
$(J_{1-} - J_{2+})|\psi > = 0$.
The solution of these equations corresponds to the EPR-correlated state.
\begin{eqnarray}\label{54}
|\psi>=\frac{1}{\sqrt{N+1}}\sum_{n=0}^N |n>_N\otimes |n>_N
\end{eqnarray}
This fulfills the expression $(J_{1z} -J_{2z})|\psi> = 0$. 

The entanglement of two-qubits  is studied, they were selected from different ensembles, and then the two-qubit reduced density matrix was written by
$v=v_+=v_-$, $u$, $u^*$, $w$ elements the others $y=x_+=x^*_+=x_-=x^*_-=0$
where the basis is  $\{|00>,|01>,|10>,|11>\}$. It has a representation:
\begin{eqnarray}\label{56}
\begin{array}{rcl}
v & = & \frac{1}{4}(1\pm2<\sigma_{1z}>+<\sigma_{1z}\sigma_{2z}>),\\
w &=& \frac{1}{4}(1-<\sigma_{1z}\sigma_{2z}>)=\frac{1}{4}-\frac{<J_{1z}J_{2z}>}{N^2}\\
u&=& \frac{1}{4}(<\sigma_{1x}\sigma_{2x}>-<\sigma_{1y}\sigma_{2y}>+i2<\sigma_{1x}\sigma_{2y}>)\\
&=& \frac{<J_{1+}J_{2+}>}{N^2}.
\end{array}
\end{eqnarray}
Then the concurrence can be written as follows
\begin{eqnarray}
\begin{array}{rclcl}
C&=&2\max\{0,|u|-w\}
&=&2\max\left\{0,\frac{<J_{1+}J_{2+}>+<J_{1z}J_{2z}>}{N^2}-\frac{1}{4}\right\}.
\end{array}
\end{eqnarray}
The expectation values of $J_{1+}J_{2+}$ and $J_{1z}J_{2z}$ can be received in the state $\Psi$. 
 We get the form $C = 1/N$. The pair of the particles is an entangled state in the manyfold. 
 
 \section{Kicked Top model}\label{ktm}

The quantum kicked top(QKT) model is a finite-dimensional dynamical system that is studied by quantum chaos for its compact phase space and the formation of chaoticity structure depends on the  $\kappa$ parameter \cite{7fhmkrs}.
The quantum kicked top is a combination of a rotation and a torsion, i.e. it describes a spin of size $j$ precessing about the $y$-axis together with a state-dependent twist about the $z$-axis which is characterized by the driving parameter $\kappa$. The period between kicks is $\tau$ and $p$ means the amount of the $y$-precession within one period. This model is governed by the Hamiltonian \cite{fh2}
\begin{eqnarray}
H=\frac{\kappa_0}{2j}J_z^2 \sum_{n=-\infty}^{\infty}\delta (t-n\tau)+\frac{p}{\tau}J_y
\end{eqnarray}
Here $J_{x,y,z}$ are generators of the angular momentum
operator {\bf J} where $J_i,J_j=i\varepsilon_{ijk}J_k$. The Floquet map is the unitary operator:
\begin{eqnarray}\label{qktu}
U = \exp[-i(\kappa_0/2j\hslash )J_z^2]\exp[-i(p/\hslash )J_y]
\end{eqnarray}
which evolves from states after a kick to the
next. 
The constants take the values $\hslash = 1$ and $p=\pi/2$.
The $\kappa_0$ parameter, which is the amount of twist applied between kicks, controls the transition and extent of the chaos. 
When it disappears, the dynamic is a rotation. 
Since the magnitude of the total angular momentum is conserved, the quantum number $j$, where the eigenvalues of ${\bf J}^2$ are $j(j + 1)\hslash^2$.
 These show what can be produced as a regular or mixed-phase space, the amount of chaotic orbits is negligibly small at 0. In the case of $\kappa_0 = 0$, the classical map can be integrated, a rotation, but if $\kappa_0 > 0$ then chaotic orbits appear in the phase space, and when $\kappa_0 > \kappa_{cr}$, the system becomes completely chaotic.
Corresponding to the many-body model is possible for consideration
the large $\bf J$ spin as the total spin of the spin=1/2 qubits, substituting $J_{x,y,z}$ with $\sum_{l=1}^{2j}\sigma_l^{x, y, z}/$2 \cite{8xwsgbcsbh}. 
Therefore the Floquet operator can  be described by the $2j$ qubit, which corresponds to an Ising model with fully homogeneous coupling and a transverse magnetic field:
\begin{eqnarray}
U=\exp \left(-i\frac{\kappa_0}{4j}\sum_{l<l'=1}^{2j}\sigma_l^{z}\sigma_{l'}^z)\right)\exp\left(-i\frac{\pi}{4}\sum_{l=1}^{2j}\sigma_l^y\right)
\end{eqnarray}
where $\sigma_l^{x,y.z}$ are the Pauli matrices. Usually, the $2j + 1$ dimensional permutation symmetric subspace of the  $2^{2j}$ dimensional full space corresponds to the kicked top system.
Since the multiples of $2\pi j$ are $\kappa_0$, $U$ is a local operator and does not produce entanglement, so the interval of calculations is chosen
 $\kappa_0 \in [0, \pi j]$. But they investigated the case of 2-qubit $j = 1$ \cite{rlp2}, where they studied systems that are not related to the classical limit and in this connection, several quantum correlations have also been determined in \cite{bs}. 

In the case of $j = 3/2$, the three-qubit case is examined, i.e. each with only the nearest-neighbor periodic boundary conditions. This so-called nearest-neighbor excitation is a transverse Ising model \cite{51tp,52alvs}.
The superconducting Josephson junction experiment can be considered in the \cite{nrfck}, which shows chaoticity.

\subsection{Three-Qubit case}

The Equation (\ref{qktu}), the unitary Floquet operator for $2j = 3$-qubits, which describes the dynamics of a 3/2 spin during a kicked Hamiltonian following
\begin{eqnarray}
U=\exp\left(-i\frac{\kappa_0}{6}(\sigma_1^z \sigma_2^z+\sigma_2^z\sigma_3^z+\sigma_3^z\sigma_1^z)) \exp(-i\frac{\pi}{4}(\sigma_1^y+\sigma_2^y+\sigma_3^y)\right)
\end{eqnarray}
where all terms have the meaning given in \cite{sdvmal}.
The solution of the 3-qubit case starts from the fact that
$[U,\otimes_{l=1}^{2j}\sigma_l^y]=0,$
 there is up-down or parity symmetry.
The 4-dimensional spin-quartet permutation symmetric space $j = 3/2,$ $\{|000>,|W>=(|001>+|010>+|100>)/\sqrt{3}, |\overline {W}>=(|110>+|101>+|011>)/\sqrt{3}, |111>\}$ shows the parity symmetry to express the basis
$|\Phi_1^{\pm}>=\frac{1}{\sqrt{2}}(|000>\mp i|111>)$, and
$|\Phi_2^{\pm}>=\frac{1}{\sqrt{2}}(|W>\pm i|\overline{W}>)$.

These are parity eigenstates which $\otimes_{l=1}^3\sigma_l^y|\Phi_j^{\pm}>=\pm |\Phi_j^{\pm}>$.
In this system, $|W>$ means the quantum information state and $|\Phi_1^{\pm}>$ fulfills the GHZ states.
While the GHZ class of states is particularly situated at the poles of the sphere, the superposition of the $W$ states is on the equatorial plane and the peaks at $(\Theta_0=\pi/2, \Phi=\pm \pi/ 2) $.
These points fulfill the low-order periodic points of the classical map and represent the initial states of the quantum system.
The unitary operator $U$ is introduced  by the following expression
$u_{11}=U_+$, $u_{22}=U_-$ and $u_{12}=u_{21}=0$
 where 0 is a $2\times 2$ null matrix, and the $2\times 2$ dimensional blocks  matrix $U_+ (U_-)$ are given on the bases  $\{\Phi_1^+,\Phi_2^+\}(\{ \Phi_1^-,\Phi_2^-\})$, they are found in the subspaces with positive (negative) parity.
The matrix elements are determined:
\begin{eqnarray}\label{qktu1}
U_{\pm}=\pm e^{\mp\frac{i\pi}{4}}e^{-i\kappa}\left(\begin{array}{cc}
\frac{1}{2}e^{-2i\kappa} & \mp\frac{\sqrt{3}}{2}e^{-2i\kappa}\\
\pm \frac{\sqrt{3}}{2}e^{2i\kappa} & -\frac{i}{2}e^{2i\kappa}
\end{array}\right)
\end{eqnarray}
 The parameter $\kappa=\kappa_0/6$ are chosen.
 The form $U_+$ means a rotation $e^{-i\gamma \overrightarrow{\sigma}\overrightarrow{\eta}}$ by angle $\gamma$ about an axis $\hat{\eta}=\sin \Theta \cos \Phi \hat{x}+\sin \Theta \sin \Phi \hat{y}+\cos \Theta \hat{z}$ to a phase.
 Then the association to the equation (\ref{qktu1})  can be written as $\cos \gamma =\frac{1}{2} \sin 2 \kappa$, $\Phi=\pi /2+2\kappa$ and $\sin \Theta \sin \gamma = \sqrt{3}/2$.
The $U^n$ and $U_{\pm}^n$ are important considering the evolution of the initial states which is investigated:
\begin{eqnarray}
U_{\pm}^n=(\pm)^ne^{-in(\pm \frac{\pi}{4}+\kappa)}\left(\begin{array}{cc}
\alpha_n & \mp\beta_n^*\\
\pm \beta_n & \alpha_n^*
\end{array}\right)
\end{eqnarray}
where
\begin{eqnarray}
\begin{array}{rclrcl}\label{ab}
\alpha_n&=&T_n(\chi)+\frac{i}{2}U_{n-1}(\chi)\cos 2\kappa &\hspace{1.0cm}
\beta_n&=&(\sqrt{3}/2)U_{n-1}(\chi)e^{2i\kappa}.
\end{array}
\end{eqnarray}
The Chebyshev polynomials $T_n(\chi)$ and $U_{n-1}(\chi)$ are written  as follow: $T_n(\chi) = \cos(n\gamma$) and $U_{n- 1}( \chi) = \sin(n\gamma)/\sin \gamma$  where $\chi = \cos \gamma = sin(2\kappa)/2$ and taking into account that $|\alpha_n |^2+|\beta_n |^2=1$.
\begin{eqnarray}
T_n^2(x)+(1-x^2)U_{n-1}^2(x)=1
\end{eqnarray}
This was considered that the Pell identity is fulfilled by Chebyshev polynomials through the unitarity of quantum mechanics \cite{sdvmal}.
The interval of $\chi$ is limited in this case   $|\chi|\le 1/2$ where the  expression contents
$|T_n(\chi)|\le 1$, and from the equation (\ref{ab}) goes on  $|U_{n-1}|\le 2/\sqrt{3}$.
The time evolution of a three-qubit permutation was considered with a symmetric state and then its behavior was investigated.
The three-qubit state $(|0,0>$  and $|\pi/2, -\pi/2>)$  are discussed  the hereinafter. These can be applied to perform the time-average of concurrence between any two qubits characterizing the entanglement.
In the case of the semiclassical limit in phase space 
a three-qubit state 
$\otimes^3|0>$ fulfills  a coherent state at $|0,0>$ which corresponds to the period-4 trajectory on classical system. The $\otimes^3|+>_y$ satisfies the coherent state at $|\pi/2,-\pi/2>$, which is a suitable fixed point on the classical phase space limit.  Due to the parameter $\kappa$ changing this is unstable and goes from regular to chaotic phase space through $\kappa_0=\kappa_{cr}$ \cite{af}.

\paragraph{Initial state $|000> = |\Theta_0=0, \Phi=0>$}

The state on the period-4 trajectory fulfills the coherent state at $|000>=\otimes^3|0>$ following
\begin{eqnarray}
\begin{array}{rcl}
|\Psi_n>=U^n|000> &=&\frac{1}{\sqrt{2}}U^n(|\Phi_1^+>+|\Phi_1^->)\\
&=&\frac{1}{\sqrt{2}}(U^n_+(|\Phi_1^+>+U^n_-(|\Phi_1^->)\\
&=&\frac{1}{2}e^{-in(\frac{3\pi}{4}+\kappa)}\{(1+i^n)(\alpha_n|000>+i\beta_n|\overline{W}>)\\
&+&(1-i^n)(i\alpha_n|111>-\beta_n|W>)\}
\end{array}
\end{eqnarray}
The reduced density matrices $\rho_1(n) = tr_{2,3}(|\Psi_n><\Psi_n|)$, $\rho_{12}(n) = tr_3(|\Psi_n><\Psi_n|)$ are obtained by the  1 and 2 qubit. 
The 2-qubit reduced matrix arises from one-qubit with the two others. 
 The concurrence is able to be expressed by the entanglement between two qubits \cite{40wkw}.

\paragraph{Concurrence}

We introduced the entanglement between any two qubits and this was quantified by the concurrence (\ref{ent}) in section (\ref{con}).
It doesn't matter which two qubits are chosen due to state permutation symmetry, there exists only one concurrence.
 The concurrence which was raised from the two-qubit reduced density matrix that was introduced by the difference of the square roots of the eigenvalues 
in decreasing order of  $(\sigma_y \bigotimes \sigma_y)\rho_{12}(\sigma_y \bigotimes \sigma_y) \rho^*_{12}$ and $\rho^*_{12}$ is conjugation is on $(\sigma_z)$ basis.

This expression between any two
qubits in the state $|\Psi_n>$ of equation ({\ref{con}})
can be written as the two-qubit state is an $X$ state \cite{57tyjhe}
 when the time $n$ is even. A two-qubit reduced density operator
of $\rho_{12}(n)$ is received by  tracing out one of the qubits in
$|\Psi_n><\Psi_n|$ which is given:
\begin{eqnarray}
\rho_{12}(n)=\left(
\begin{array}{cccc}
|\alpha_n|^2 & 0 & 0 & -\frac{i}{\sqrt{3}}\alpha_n\beta_n^* \\
0 & \frac{1}{3}|\beta_n|^2 & \frac{1}{3}|\beta_n|^2 & 0\\
0 & \frac{1}{3}|\beta_n|^2 & \frac{1}{3}|\beta_n|^2 & 0\\
\frac{1}{\sqrt{3}}\alpha_n^*\beta_n & 0 &0& \frac{1}{3}|\beta_n|^2
\end{array}
\right)
\end{eqnarray}
 whose concurrence is obtained from the expression considering
the $X$ states \cite{57tyjhe},
\begin{eqnarray}
\begin{array}{rcl}
C(n,\kappa_0)&=& 2\max\left[0,\frac{1}{3}|\beta_n|^2-\frac{1}{\sqrt{3}}|\alpha_n||\beta_n|,-(\frac{1}{3}|\beta_n|^2-\frac{1}{\sqrt{3}}|\alpha_n||\beta_n|)\right]\\
&=&|U_{n-1}(\chi)|\left|\frac{1}{2}||U_{n-1}(\chi)|-\sqrt{1-\frac{3}{4}|U_{n-1}(\chi)|^2}\right|
\end{array}
\end{eqnarray}
where  $\chi= \cos \gamma=  sin(2\kappa)/2 = sin(\kappa_0/3)/2$ was applied. 
This is reliable when the time $n$ is even i.e. 
$C(2m -1, \kappa_0) = C(2m, \kappa_0)\;\;\; m = 1, 2\dots$
This expression shows how much concurrence is produced in the first step:
$C(1,\kappa_0)=\sin(\kappa_0/3)\left[(1-\frac{3}{4}\sin^2(\kappa_0/3))^{1/2}-\frac{1}{2}\sin (\kappa_0/3)\right]$
which is valid when $0\le \kappa_0 \le 3\pi$, and beyond this interval the
concurrence becomes periodic. 
This shows monotonic behavior
from 0 to $\kappa_0= \pi/2$ where it reaches a maximum and its value is $(\sqrt{13}-1)/8$.

\paragraph{Analysis}

\begin{figure}
\begin{center}
\includegraphics[width=5.0cm]{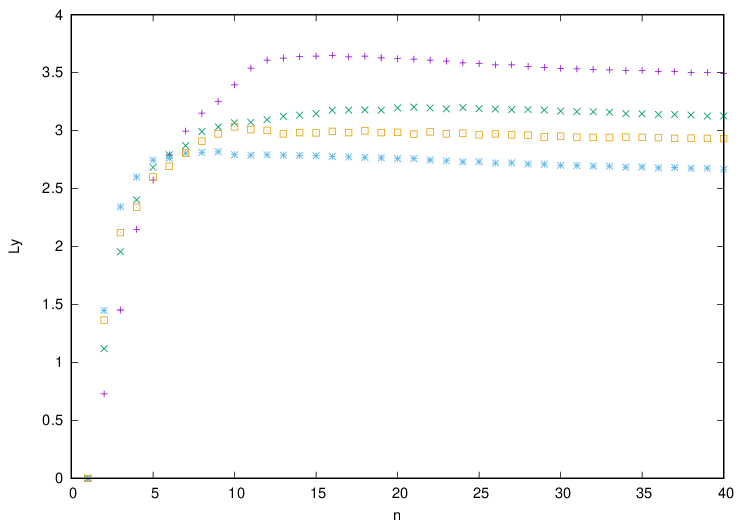}
\caption{ The Lyapunov exponent of Kicked top model depends on the time $n\tau$. This quantity as a function strength of the driven  $\kappa=0.2,0.4,0.6$ and $1.002$.  }\label{fig-2}
\end{center}
\end{figure}

The entanglement and concurrence as quantum correlations can be used well to describe chaotic systems in both mixed and regular phase space.
In regular motion, the growth rate of all quantum correlations is progressive, while the initial growth rate is fast in mixed and chaotic space.
The quick growth refers to classical Lyapunov exponents it is seen in the Figure (\ref{fig-2}), where the driven parameter $\kappa=0.2$.
 
\begin{figure}
\begin{center}
\includegraphics[width=5.0cm]{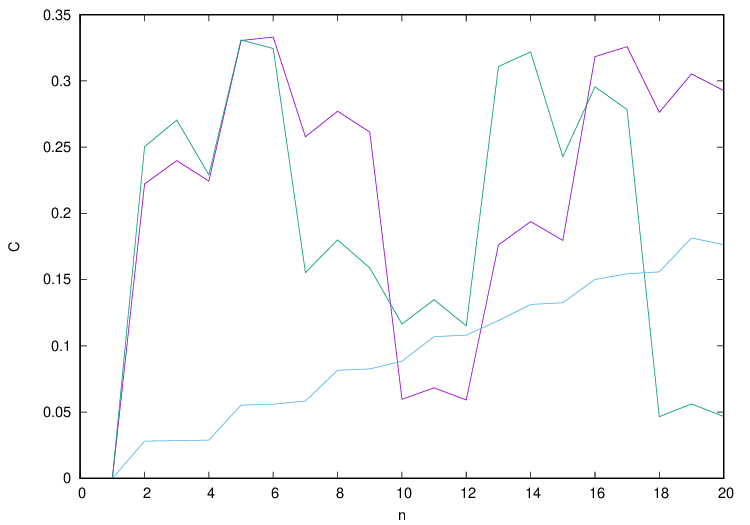}
\includegraphics[width=5.0cm]{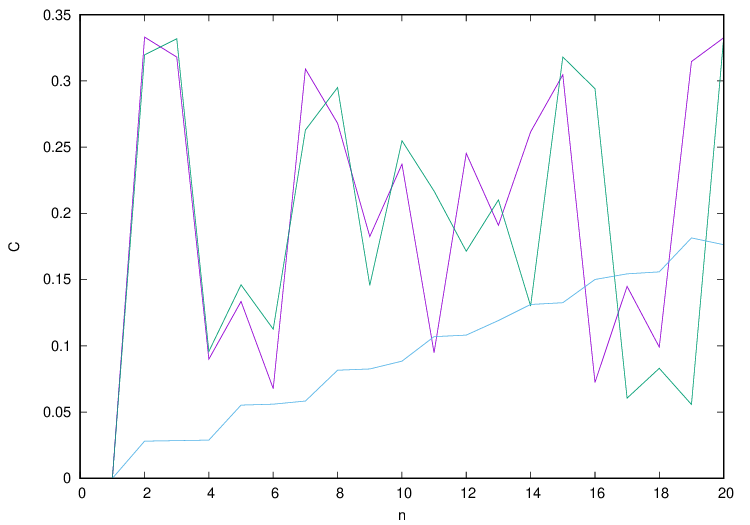}
\caption{Concurrence of Kicked Top model showes the intermittent depending on $n$, where the parameter $\kappa_0 = 0.6, 0.7$ and $0.2$ }\label{fig-4}
\end{center}
\end{figure}

At small values of the chaotic parameter $\kappa$, the entanglement dynamics corresponds to a good approximation to the semiclassical description. 
Therefore, the degree of chaos in the system affects the creation of correlations through the time-dependent model.
The simulation of the system dynamics shows an intermittent behavior of the correlations (Figure \ref{fig-4}), which is confirmed in the experiment\cite{nrfck}.
This is because at these points the dynamics become local and do not change the value of the correlations.

\begin{figure}
\begin{center}
\includegraphics[width=5.0cm]{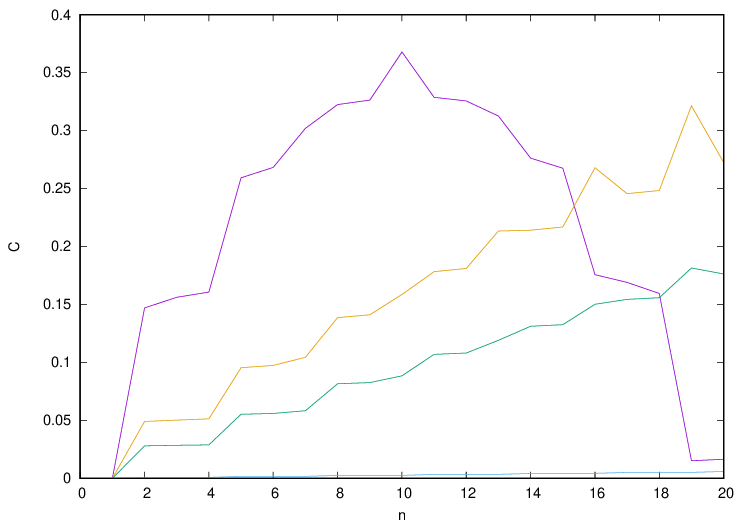}
\caption{Concurrence of Kicked Top model depending on $n$ is
plotted, where $\kappa=0.2, 0.4$ and $1.002$.}\label{fig-3}
\end{center}
\end{figure}

As a result of the simulations the behavior of the correlations indicates the following dynamics.
Beginning, there are uncorrelated states, then the behavior of the quantum correlation curve continues as it rises.
The concurrence increases initially, but as the states of the two qubits become more and more mixed, the concurrence starts to decrease and becomes zero (Figure \ref {fig-3}). 

In the multi-particle system, the entanglement records the distribution of the correlations in several particles, that is the bipartite correlation according to concurrence (Section \ref{dm}) transforms into multipart-correlations.
 This is the state when the entanglement began to be shared globally, not bipartitely.
 In this case, the entanglement is divided into several parts.

Chaos arises due to the connection of degrees of freedom.
In quantum systems, the relation of degrees of freedom indicates the formation of correlations.

To describe quantum correlations, pseudorandom states are suitable for providing superpositions and characterizing chaotic dynamics.

The problem of quantum-classical transitions and the role of quantum correlations have great significance in these systems.
Therefore, Monte-Carlo methods can also be used well in such cases.

We note that classical chaos is characterized by uncertainty and Lyapunov exponents, which in the case of quantum chaos is realized in the superposition of the quantum states of a quantum system.
A quantum trajectory follows the classical orbit, its conditions can be seen based on a given external observable.
It was shown that classical trajectories can be reconstructed from quantum systems if the quantum system is continuously measured \cite{tbshkj} with appropriate measurement strength.
If the measuring signal strength is significant enough to localize the wave packet to such an extent that the effect of interference belonging to different {\em paths} does not have a great significance, it can be neglected to keep the feedback noise at a low value, the system follows the classical paths.
It has been investigated that the entanglement between the two subsystems shows a strong feedback effect, and that classical paths do not form in this regime.
To develop classical tracks, both subsystems would be classical enough.

That is Ehrenfest {\em break time} \cite{fccp} which means the time when the quantum expectation values and the classical motion start to behave differently compared to each other, which is exponentially small in chaotic systems compared to normal systems.
In other words, quantum chaotic multipart systems represent an important research area in the investigation of the interaction of chaos, quantum correlations and measurements, as well as their role in the quantum-classical transition.

 The kicked top model was studied in this article. This system has great significance in the research area of quantum chaos both theoretically and experimentally.
In the multi-qubit Hilbert space, the initial coherent state changes from the regular classical phase space to the chaotic region, the long-time averaged  of the correlations reflects the behavior of the system, which can be represented by a classical stroboscopic map \cite{af}.
Note that entanglement  appears
even though the initial state is localized on a stable island, there is no chaos in the phase space.
Classical bifurcations, ergodicity \cite{34vmsdal} and the relationship between phase transformation \cite{jbpszaf} in phase space have significance in the research of entanglement within dynamic systems.

\section{Conclusion}

In this article, we considered the concurrence in a bipartite system with the pairwise method in the case of spin-coherent states, Dicke states and EPR-correlated ensembles. The approximation of QKT model was discussed on the three-qubit by the solvable model. The chaoticity appears in this system.

Bi-partite entanglement can be 
expressed by the pairwise entanglement satisfying the symmetry condition.
The two-particle density matrix was introduced by the expectation values of the collective operator in the case of symmetrical states.
The pairwise method is suited for the bigger ensemble, provided that it fulfills the symmetry criterion.

This resulted in the spin coherent states don't have entanglement because the concurrence vanishes. The Dicke states contain a higher level of entanglement. The concurrence is plotted as a function $M$ for different $N$ on the Figure (\ref{fig-1}). In the case of an EPR-correlated ensemble the concurrence $C=1/N$ due to the pair of the particles fulfilling an entangled state in the manyfold.

We note that the significant entanglement disappears when the multi-particle state is partitioned in pairs i.e. the multiparticle state is split into pairs, which can cause problems during quantum computing.
The entanglement of formation becomes more difficult in many-particle systems for mixed states with dimensions larger than 2.
Studying the two-particle concurrence with many-body systems can be a useful way to consider the more complicated problem.

The QKT model is widely used to investigate the description of quantum chaos.
Because entanglement and chaotic behavior are of great significance in the operation of a quantum computer to minimize errors. Therefore it is important to get to know these phenomena as thoroughly as possible. In this article, we investigated the entanglement and concurrence of the QKT model.

Since the QKT model contains periodic excitation, it is therefore highly nonlinear. We used a 3-qubit description within the given range of the $\kappa$ parameter, where theoretically an accurate description can be given.

The exactly solvable 3-qubit instances
the kicked top model provides insight into how entanglement thermalizes in closed quantum systems a the perception of long-time averages approaching ensemble averages, as the classical limit approaches global chaos, as
predicted by random matrix theory.
Thus, it became possible to study the examination of the dynamic properties of the model for 3 qubits and control the dependence on the value of the excitation parameter.

Additional numerical approximations must be performed to know the behavior of the model with greater accuracy over a wider range of the parameters on a larger many-body system.
A larger number of qubits suggests a connection between non-integrability and chaos.

The dependence of the excited state of QKT on the value of the parameter $\kappa$  leads to a more precise understanding of quantum chaos.

As the quantum computer huge system which can be modeled by QKT produced errors \cite{39chbdpdjaswkw,30chb,13bs2}, where the entanglement connection to the chaos has great significance, avoiding this can reduce the chance of errors occurring.


\begin{thebibliography}{67}
\setlength{\parskip}{-3pt} 

\bibitem{ac}{ C. Adami, N.T. Cerf,}
{\em Physica D} {\bf 137} (2000) 62--69.

 \bibitem{bl1}{J. N. Bandyopadhyay and A. Lakshminarayan},
 {\em Phys. Rev. E} {\bf 69} (2004) 016201.
 
 \bibitem{jbpszaf}J. Bene, P. Szepfalusy, and A. Fulop,
{\em Phys. Rev. A} {\bf 40} 6719. 

\bibitem{39chbdpdjaswkw} { C.H. Bennett, D.P. DiVincenzo, J.A. Smolin, and W.K.
Wootters,} 
{\em Phys. Rev. A} {\bf 54}(1996)  3824. 

 \bibitem{30chb} C.H. Bennett et al.,  
  {\em Phys Rev Lett} {\bf 76} (1996) 722. 

\bibitem{12gbgcdls}G. Benenti, G. Casati, and D.L. Shepelyansky,
{\em  Eur. Phys. J. D} {\bf 17} (2001)265. 


\bibitem{13gpbfbfmivit} G.P. Berman, F. Borgonovi, F.M. Izrailev, and V.I.
Tsifrinovich,
{\em  Phys. Rev. E} {\bf 64} (2001) 156226.


\bibitem{34sbdls} S. Bettelli and D.L. Shepelyansky, 
{\em Phys. Rev. A} {\bf 67}(2003)054303. 

\bibitem{tbshkj}T. Bhattacharya, S. Habib, and K. Jacobs,
{\em Phys. Rev. A} {\bf 67}(2003)042103. 

 \bibitem{bs}{U. T. Bhosale and M. S. Santhanam},
{\em Phys. Rev. E} {\bf 98} (2018) 052228. 

  \bibitem{13bs2}U. T. Bhosale and M. S. Santhanam,
  {\em Phys. Rev. E} {\bf 95} (2017) 012216.
  

\bibitem{44db} D. Bruß,
 {\em J. Math. Phys}. {\bf 43} (2002)4237.

\bibitem{7ggcgbgc} G.G. Carlo, G. Benenti, and G. Casati, 
{\em Phys. Rev. Lett.} {\bf 91}(2003) 257903.

\bibitem{7a} G.G. Carlo, G. Benenti, G. Casati, and C. Mejıa-Monasterio, 
{\em Phys. Rev. A} {\bf 69} (2004) 062317. 

\bibitem{fccp} F. Cametti1 and C. Presilla1
Phys. Rev. Lett. {\bf 89} (2002) 040403.

 \bibitem{csagj} S. Chaudhury, A. Smith, B. E. Anderson, S. Ghose, and P. S.
Jessen,
 {\em Nature} {\bf 461} (2009) 768. 

\bibitem{sdvmal}{ S. Dogra, V. Madhok, A. Lakshminarayan,}
{\em Phys. Rev E} {\bf 99} (2019) 062217.

\bibitem{1wdgvjic}W. D\"ur, G. Vidal, J.I. Cirac,
 {\em Phys. Rev. A} {\bf 62},(2000) 062314. 

\bibitem{af}{ A. Fulop,}
{\em Acta Univ. Sapientiae Informatica} {\bf 12} 2 (2020) 283--301.

\bibitem{11vvf} V.V. Flambaum, 
 {\em Aust. J. Phys.} {\bf 53} (2000)489.

\bibitem{fmt1}H. Fujisaki, T. Miyadera, and A. Tanaka,
{\em Phys. Rev. E} {\bf 67}, (2003)066201.
 

\bibitem{10bgdls} B. Georgeot and D.L. Shepelyansky, 
{\em Phys. Rev. E} {\bf 62}(2000)3504. 


\bibitem{7pgfhhw} P. Gerwinski, F. Haake, H. Wiedemann,
Marek Kus, and Karol Zyczkowski,
{\em Phys. Rev. Lett.} {\bf 74} (1995)1562–1565.

\bibitem{10sgbcs} S. Ghose and B. C. Sanders,
  {\em Phys. Rev. A} {\bf 70}(2004)062315.
  
\bibitem{7fhmkrs} F Haake, M Kus, R Scharf, 
{\em Zeitschrift f\"ur Physik B Condensed Matter} {\bf 65}(1987)381–395.
  
 \bibitem{fh2} {F. Haake}, Quantum Signatures of Chaos, Spring-Verlag,
Berlin, 1991. 

\bibitem{mkrsfh4} M Kus, R Scharf, and F Haake, 
{\em Zeitschrift f\"ur Physik B Condensed Matter} {\bf 66}(1) (1987)129–134.

\bibitem{5mkjmfh} M Kus, J Mostowski, and F Haake,
{\em Journal of Physics A: Mathematical and General}
{\bf 21}(22) (1988)L1073.

 \bibitem{52alvs}A. Lakshminarayan and V. Subrahmanyam,
 {\em Phys. Rev. A} {\bf 71} (2005)062334.

\bibitem{lm2}M. Lombardi and A. Matzkin,
 {\em Phys. Rev. E} {\bf 83}, {2011} 016207.
 

\bibitem{11vmcarsg} V. Madhok, C. A. Riofrio, S. Ghose,
and Ivan H. Deutsch,
  {\em Phys. Rev. Lett.} {\bf 112} (2014)014102.
  
  \bibitem{mgtg}{V. Madhok, V. Gupta, D.A. Trottier, and S. Ghose,}
 {\em Phys. Rev. E} {\bf 91} (2015) 032906.
 
 \bibitem{34vmsdal} V. Madhok, S. Dogra, and A. Lakshminarayan,
{\em Optics Communications}, {\bf 420} (2018)189-193.

\bibitem{cmmgbggcgc}{C. Mejía-Monasterio, G. Benenti, G. G. Carlo, and G. Casati,}
{\em Phys. Rev. A} {\bf 71} (2005)062324. 

\bibitem{58skmalvs}S. K. Mishra, A. Lakshminarayan, and V. Subrahmanyam,
{\em Phys. Rev. A} {\bf 91} (2015)022318.

  
  \bibitem{4cmjpprp} C. Miquel, J.P. Paz, and R. Perazzo, 
  {\em Phys. Rev. A }{\bf 54}, (1996)2605. 

\bibitem{5cmjppwhz} C. Miquel, J.P. Paz, and W.H. Zurek, 
{\em Phys. Rev. Lett.}{\bf 78} (1997) 3971. 


\bibitem{nrfck}{C. Neill, P. Roushan, M. Fang, Y. Chen, M. Kolodrubetz, Z.
Chen, A. Megrant, R. Barends, B. Campbell, B. Chiaro et al.,}
{\em Nat. Phys.}  {\bf 12} (2016) 1037. 


\bibitem{ap3} A. Peres, Quantum Theory: Concepts and Methods,
Kluwer Academic Publishers, New York, 2002.


\bibitem{25ap} A. Peres,
 {\em Phys. Rev. Lett.} {\bf 77} (1996) 1413.

\bibitem{51tp}T. Prosen,
{\em Progress of Theoretical Physics Supplement}, {\bf 139}(191) 2000 203.

\bibitem{8drgbgc} D. Rossini, G. Benenti, and G. Casati, 
{\em Phys. Rev. A} {\bf 69}, (2004)052317.

\bibitem{rlp2}J. B. Ruebeck, J. Lin, and A. K. Pattanayak,
 {\em Phys. Rev. E} {\bf 95} (2017)062222. 
 

\bibitem{35} A.J. Scott and C.M. Caves,
 {\em J. Phys. A} {\bf 36} (2003) 9553.

\bibitem{6phsdls} P.H. Song and D.L. Shepelyansky, 
{\em Phys. Rev. Lett.} {\bf 86}(2001)2162.

\bibitem{gv} G. Vidal, 
Journal of Modern Optics 
{\bf 47} (2-3) (2000) 355.  

\bibitem{20xw}X. Wang, 
 {\em Phys. Rev. A} {\bf 64} (2001) 012313. 
 
 \bibitem{8xwsgbcsbh}X. Wang, S. Ghose, Barry C. Sanders, and B. Hu,
{\em Phys. Rev. E} {\bf 70}(2004)016217.

\bibitem{xwkm}{ X. Wang, K. Molner,}
{\em The European Physical Journal D} {\bf 18} (2002) 385--391.


\bibitem{38wkw}{W.K. Wootters,}
 {\em Quantum Inf. and Comp}. {\bf 1} (2001) 27.

\bibitem{40wkw}{ W.K. Wootters,} 
 {\em Phys. Rev. Lett.} {\bf 80} (1998)  2245. 

\bibitem{57tyjhe} {T. Yu and J. H. Eberly},
{\em  Quantum Info. Comput.}{\bf 7}(5) (2007)459-468.


\bibitem{6kz} K Zyczkowski, 
{\em Journal of Physics A: Mathematical and General} {\bf 23}(20) (1990)4427.

\end{thebibliography}
\end{document}